\pdfoutput=1

\documentclass[aps,twocolumn,showkeys,amsmath,amssymb,superscriptaddress,nofootinbib,floatfix,prc]{revtex4-1} 

\usepackage{epsfig}

\begin{document} 
\hbadness=10000

\title{Elliptic flow in ultra-relativistic collisions with light polarized nuclei}

\author{Wojciech Broniowski}
\email{Wojciech.Broniowski@ifj.edu.pl}
\affiliation{H. Niewodnicza\'nski Institute of Nuclear Physics PAN, 31-342 Cracow, Poland}
\affiliation{Institute of Physics, Jan Kochanowski University, 25-406 Kielce, Poland}

\author{Piotr Bo\.zek}
\email{Piotr.Bozek@fis.agh.edu.pl}
\affiliation{AGH University of Science and Technology, Faculty of Physics and Applied Computer Science, al. Mickiewicza 30, 30-059 Cracow, Poland}

\date{21 June 2019}  

\begin{abstract}
Estimates for elliptic flow in collisions of polarized light nuclei with spin $j\ge1$ with a heavy nucleus are presented. 
In such collisions the azimuthal symmetry is broken via polarization of the wave function of the light nucleus, resulting in 
nonzero one-body elliptic flow coefficient evaluated relative to the polarization axis. Our estimates involve experimentally 
well known features of light nuclei, such as their quadrupole moment and the charge radius, yielding 
the one-body elliptic flow coefficient in the range from 1\% for collisions with the deuteron to 5\% for for collisions with $^{10}$B nucleus. 
Prospects of addressing the issue in the upcoming fixed-target experiment at the Large Hadron Collider are discussed.
\end{abstract}

\maketitle

\section{Introduction \label{sec:intro}}

In a recent Letter~\cite{Bozek:2018xzy} we proposed a novel way to probe the collective flow formation in ultra-relativistic 
collisions of polarized deuterons with heavy nuclei, based on the measurement of the elliptic flow with 
respect to a fixed deuteron polarization axis. In the present paper we further explore this idea and extend the method to other light nuclei 
of spin $j\ge1$. We argue that nuclei with a large ratio of the quadrupole moment to the ms radius, including ${}^7$Li, ${}^9$Be, or ${}^{10}$B,
are very well suited for such studies.  

The purpose of carrying out this sort of analyses is to better understand the mechanisms standing 
behind collectivity in the so-called heavy-light systems, where the created fireball is relatively small. 
Polarized light targets offer a unique opportunity to control the 
direction of the orientation of the formed fireball, thus providing an important methodological advantage
explored in this paper: 
the elliptic flow coefficient can be measured as a one-body observable, relative to the polarization axis. 

The unexpected discovery of the near-side ridge in two-particle correlations in relative azimuth and pseudorapidity in p+A~\cite{CMS:2012qk,Abelev:2012ola,Aad:2012gla}, 
d+A~\cite{Adare:2013piz}, ${}^3$He-A~\cite{Adare:2015ctn}, and even in p+p collisions of highest 
multiplicities~\cite{Khachatryan:2010gv}, led to serious proposals that the collective behavior in small systems may have the same origin as in large fireballs formed in A-A collisions 
based on hydrodynamic or transport evolution. As a mater of fact, the early hydrodynamic predictions the elliptic and triangular flow in p+A and d+A collisions~\cite{Bozek:2011if} 
were confirmed by later experiments~\cite{CMS:2012qk,Abelev:2012ola,Aad:2012gla,Adare:2013piz,Adare:2015ctn} to a good accuracy. 

The key mechanism of the collective picture is the alleged copious
rescattering in the fireball, which leads event by event to a transmutation
of its azimuthal deformation into  harmonic flow of 
the produced hadrons. Early studies of collectivity in small systems were carried out in~\cite{Bozek:2011if,Kozlov:2014fqa,Bzdak:2014dia,Bozek:2013uha,Werner:2013tya},
followed by investigations of triangularity in $^3$He-A collisions~\cite{Nagle:2013lja,Bozek:2014cya,Bozek:2015qpa},  or studies of the $\alpha$ clusterization effects in
$^{12}$C-A collisions~\cite{Broniowski:2013dia,Bozek:2014cva,Zhang:2017xda,Zhang:2018zzu} and 
other light clustered nuclei~\cite{Rybczynski:2017nrx,Guo:2017tco,Lim:2018huo}.

The key argument speaking for the collective expansion in the evolution is the
link between the deformation of the fireball and the harmonic
flow of the produced hadrons. This deformation originates from two phenomena: random fluctuations and the ``geometry''.
Whereas in p+A  the deformation comes from fluctuations only, in d+A collisions~\cite{Bozek:2011if}
the ellipticity of the fireball is geometrically induced by the configurations of the nucleons in the deuteron, controlled by its wave function. 
This geometric effect is dominant over the fluctuations. 
In this picture, the high multiplicity events correspond to configurations where the deuteron is intrinsically oriented in the transverse plane,
when its two nucleons are transversely well separated. This yields a large number of participant nucleons from the other nucleus, and simultaneously 
a large elliptic deformation~\cite{Bozek:2011if}.
A generalization of this argument holds also for the case of triangular deformation in ${}^3$He~\cite{Nagle:2013lja,Broniowski:2013dia,Bozek:2014cya}. 
The experimental analysis carried out by the PHENIX Collaboration confirms this scenario in the found
hierarchy of the elliptic and triangular flows measured in p+Au, d+Au, and $^3$He+Au collisions~\cite{Adare:2013piz,Adare:2015ctn,Aidala:2018mcw},
in support of the outlined mechanism of collectivity formation.

An entirely different point of view on the small systems is promoted in the studies within the Color Glass Condensate (CGC) theory 
(for reviews and references see, e.g.,~\cite{McLerran:2001sr,Gelis:2012ri}).
There, the correlations are dominantly generated in the earliest phase of the collision via the coherent gluon 
production~\cite{Dumitru:2008wn,Dusling:2012iga,Kovchegov:2012nd,Dusling:2013oia}.
In the case of d+A collisions, one would then expect that configurations corresponding to highest multiplicity events have color domains 
localized in the transverse plane around the two largely separated nucleons from the deuteron. Since these two color 
domains would contribute independently, forming a ``double'' p+A 
collision, the elliptic flow in d+A would be smaller than in p+A collisions, unlike in the experiment. 
The argument outlined above, however, was recently circumvented in~\cite{Mace:2018vwq,Mace:2018yvl,Mace:2019rtt}, where contrary to naive expectations the highest 
multiplicity events in CGC correspond to configurations where the nucleons from the deuteron are one behind the other, which leads to larger saturation scales.
See also the discussion in~\cite{Nagle:2018ybc}.

The method and the experimental proposal proposed in~\cite{Bozek:2018xzy} and further corroborated here allows us to control, to some extent, the orientation 
of the small nucleus, and thus the orientation of the fireball. It may thus help, by constraining the geometry, to resolve the fundamental question if the angular correlations in small systems 
originate from the early stage dynamics of glue or from the later interactions in the fireball. 

\section{Basic idea \label{sec:basic}}

Several light nuclei, such as the deuteron, ${}^7$Li, ${}^9$Be, or ${}^{10}$B, possess angular momentum $j$, hence the states of good $j_3$ 
have non-zero magnetic moment and can be polarized. When the corresponding wave function contains orbital angular momentum components
with orbital angular momentum $L>0$ , the distribution of the nucleons  
in states of good $j_3$ is not spherically symmetric. Thus polarization controls to some 
degree the ``shape'' of the nuclear distribution in the collision.

\begin{figure}
\centering
\includegraphics[angle=0,width=0.2 \textwidth]{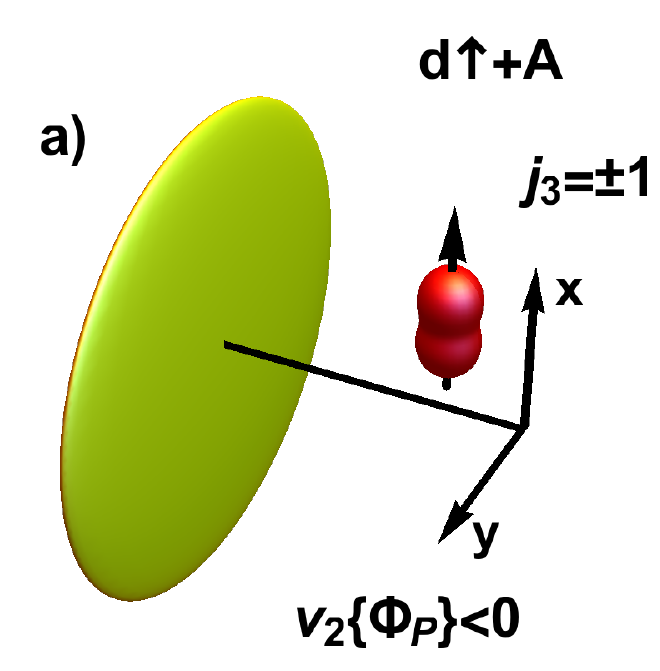} \hfill
\includegraphics[angle=0,width=0.2 \textwidth]{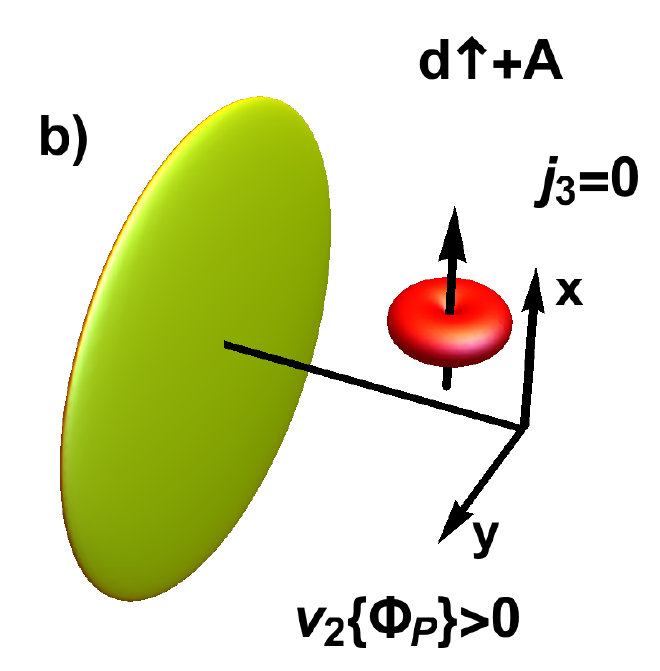} 
\vspace{-1mm}
\caption{A cartoon of the ultra-relativistic collision of a heavy nucleus on a polarized deuteron target. The deuteron is polarized along the direction $\Phi_P$ perpendicular to the beam axis and has
the angular momentum projection $j_3=\pm 1$ (panel a) or $j_3=0$ (panel b). 
The orientation of the created fireball reflects the deformation of the polarized deuteron distribution, which is prolate in (a) and oblate in (b). 
The shape-flow transmutation during the collective evolution yields to the elliptic flow 
coefficient $v_2\{\Phi_P\}$ with the signs as given in the figure. \label{fig:polar}}
\end{figure}

The case of the deuteron is presented in Fig.~\ref{fig:polar}. The polarization axis, denoted as $\Phi_P$ (which is the angular momentum quantization axis in the deuteron's rest frame),
is chosen perpendicular to the beam. The situation from panel (a) correspond to the deuteron with
angular momentum projection $j_3=\pm 1$, which leads to a prolate probability distribution of the nucleons in the deuteron, as well as, correspondingly, to 
a prolate fireball stretched along $\Phi_P$. If the collectivity mechanism is based on the shape-flow transmutation, then the 
resulting elliptic flow coefficient evaluated in reference to $\Phi_P$ (defined in Sec.~\ref{sec:elli}) is negative, $v_2\{\Phi_P\}<0$.
Contrary, for $j_3=0$  shown in panel (b), we find an oblate shape and $v_2\{\Phi_P\}>0$. 

\begin{figure}
\centering
\includegraphics[angle=0,width=0.2 \textwidth]{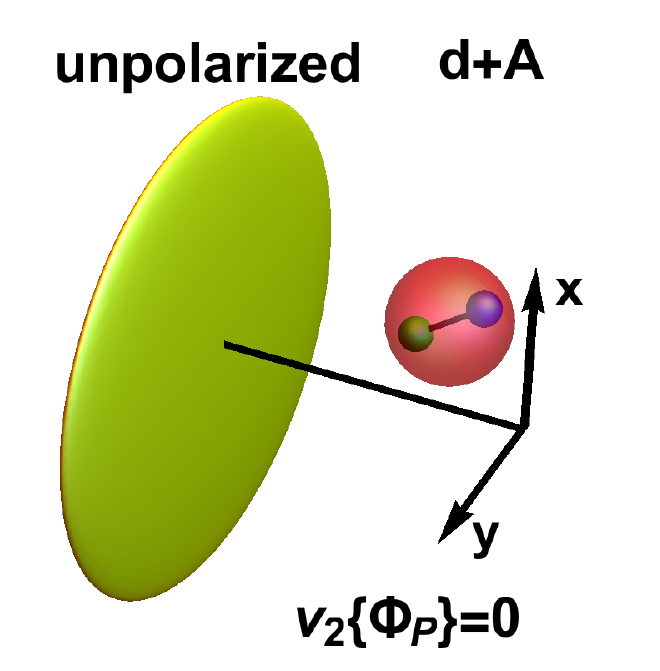} 
\vspace{-1mm}
\caption{Same as in Fig.~\ref{fig:polar}, but for the case of unpolarized deuteron  \label{fig:unpolar}}
\end{figure}

For the case of the unpolarized deuteron (Fig.~\ref{fig:unpolar}) the distribution of the nucleons in the deuteron and, consequently, the shape of the formed fireball is (up to fluctuations) 
azimuthally symmetric, hence 
for any fixed axis $\Phi_P$ the one-body flow coefficient averages to zero,  $v_2\{\Phi_P\}=0$.

\section{Ellipticity with respect to the polarization axis \label{sec:elli}}

\subsection{Generic definition of eccentricity \label{sec:generic}}

Let us define in the usual way the eccentricity vector of rank $n$ for a general distribution of sources in the transverse plane
{\em in a given event}, $f(\vec{\rho})$,  as
\begin{eqnarray}
 \vec{\epsilon}_n &=& (\epsilon^x_n, \epsilon^y_n), \;\;\; \epsilon^x_n + i  \epsilon^y_n  
 = - \frac{\int d^2\rho \,  e^{i n \alpha}\rho^2 f(\vec{\rho})}{ \int d^2\rho  \, \rho^2 f(\vec{\rho})}. \label{eq:eps} 
\end{eqnarray}
Here $\rho=\sqrt{x^2+y^2}$ is the transverse coordinate and $\alpha={\rm arctan}(y/x)$ is the azimuth. The overall sign is conventional
and chosen in such a way that the signs of 
the eccentricity of the initial fireball and the corresponding harmonic flow coefficient, $v_n$, are the same.
It is understood that the calculation is made in the center-of-mass frame, where $ \int d^2\rho \, \vec{\rho} f(\vec{\rho})=0$,

Assuming a reference system where the polarization axis $\Phi_P$ is along the $x$ axis (cf. Fig.~\ref{fig:polar}),
we need the projection of ellipticity on $\Phi_P$, i.e., the $x$ component of Eq.~(\ref{eq:eps}),
\begin{eqnarray}
\epsilon^x_2 = -  \frac{\int d^2\rho \, (x^2-y^2) f(\vec{\rho})}{ \int d^2\rho  \, (x^2+y^2) f(\vec{\rho})}. \label{eq:epsphiP}
\end{eqnarray}
Taking for simplicity point-like sources, in which case $f(\vec{\rho})=\sum_i \delta(x_i-x) \delta(y_i-y)$, we can write 
(in each event)
\begin{eqnarray}
\epsilon_2^x = -  \frac{\sum_{i=1}^N (x_i^2-y_i^2) }{ \sum_{i=1}^N  (x_i^2+y_i^2)}, \label{eq:epsphiPs}
\end{eqnarray}
where $N$ is the number of sources.
Next, we need to average Eq.~(\ref{eq:epsphiPs}) over events, denoted with brackets, $\langle . \rangle$, to get the 
ellipticity of the fireball with respect to the polarization axis $\Phi_P$,
\begin{eqnarray}
\epsilon_2\{\Phi_P\} \equiv - \left \langle  \frac{\sum_{i=1}^N (x_i^2-y_i^2) }{ \sum_{i=1}^N  (x_i^2+y_i^2)} \right \rangle . \label{eq:epsphiPav}
\end{eqnarray}
This is how the ellipticity is evaluated, in particular, in Monte Carlo simulations.

When the number of sources is large, one may approximate the average of ratios by the ratio of the averages as follows,
\begin{eqnarray}
\epsilon_2\{\Phi_P\} &=& -  \frac{\left \langle \sum_{i=1}^N (x_i^2-y_i^2)  \right \rangle}{\left \langle \sum_{i=1}^N  (x_i^2+y_i^2) \right \rangle} 
+ {\cal O}(\tfrac{1}{N}) \nonumber \\ &=&  -  \frac{\left \langle x^2-y^2  \right \rangle}{\left \langle x^2+y^2 \right \rangle} 
+ {\cal O}(\tfrac{1}{N}). \label{eq:ratio3}
\end{eqnarray}
This expression will be useful for the estimates made in the following sections.

\subsection{From ellipticity of the nuclear distribution to ellipticity of the fireball \label{sec:nuclfire}}

\begin{figure}
\vspace{-10mm}
\centering
\includegraphics[angle=0,width=0.27 \textwidth]{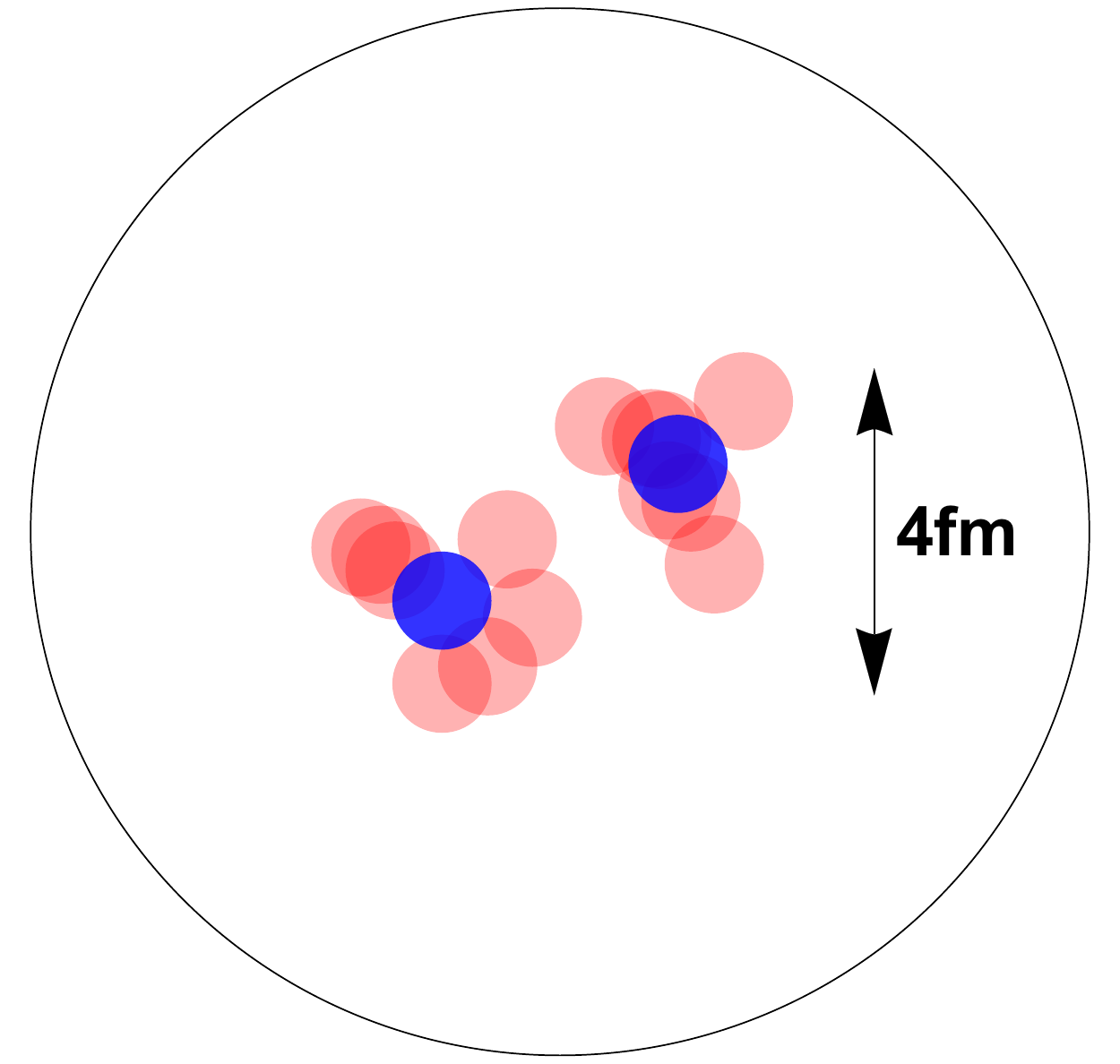}
\vspace{-2mm}
\caption{A transverse section through the fireball formed a sample d+Pb event in the Glauber Monte Carlo simulation. 
The dark disks indicate the transverse positions of the nucleons in the deuteron, and the 
light disks are the wounded nucleons from Pb, which quench the ellipticity. Note the typical 
large separation between the nucleons in the deuteron, corresponding to the large 
deuteron size. The outer circle indicates the size of Pb nucleus. \label{fig:event}}
\end{figure}

In a collision of a light projectile on a heavy target all the nucleons from the projectile participate in the collision, except for very peripheral collisions. 
In the Glauber model, for central (high multiplicity) events in a light-heavy collisions system one selects exclusively events  where all the nucleons from the small projectile participate. The impact of the  deformed nucleon distribution in the small projectile on the large uniform density of the target creates a fireball with a similar deformation  \cite{Bozek:2011if}.
To get a first estimate of the size of the deformation in the whole fireball
one can  calculate the ellipticity using the positions of the  positions of nucleons in the small projectile.
With Reid93 deuteron wave functions Eq.~(\ref{eq:epsphiPav}) yields
\begin{eqnarray}
\epsilon_2^{|\Psi|^2_{j_3=0}}\{\Phi_P\} \simeq 0.14 \ , \;\; \epsilon_2^{|\Psi|^2_{j_3=1}}\{\Phi_P\} \simeq -0.07 \ .  \label{eq:deutnum}
\end{eqnarray}
The washing out of the distribution of the nucleons in the deuteron by the wounded nucleons from Pb is illustrated in Fig.~\ref{fig:event}.  
Additional participants from the large nucleus reduce slightly the elliptic deformation.

We may estimate the size magnitude of the washing-out effect from the knowledge of the 
wounding distance between the nucleons. A nucleon from one projectile interacts with a nucleon from the other projectile when their impact factor $\vec{b}$ is sufficiently small. 
The collision occurs with the probability $P_{\rm in}(b)$, with $\int \!\! d^2 b\, P_{\rm in}(b)=\sigma_{\rm in}$, the $NN$ inelastic cross section.
We may thus consider overlaying the distribution of $\vec{b}$ over 
the positions of the nucleons from the lighter projectile. Then (for central collisions) the dispersions of the distributions are changed 
into $\langle x^2 \rangle \to  \langle x^2 \rangle +  \langle b^2 \rangle/2$, $\langle y^2 \rangle \to  \langle y^2 \rangle +  \langle b^2 \rangle/2$.
As a result, the numerator of Eq.~(\ref{eq:ratio3}) is unchanged, whereas the denominator is increased by $ \langle b^2 \rangle$. In consequence, the 
quenching factor between the ellipticity of the light projectile distribution, $\epsilon^{|\Psi|^2}_2\{\Phi_P\}$, and the ellipticity of the fireball, $\epsilon_2\{\Phi_P\}$, is approximately
\begin{eqnarray}
\epsilon_2\{\Phi_P\} \simeq \frac{\tfrac{2}{3} \langle r^2 \rangle}{\tfrac{2}{3} \langle r^2 \rangle+ \langle b^2 \rangle} \epsilon^{|\Psi|^2}_2\{\Phi_P\}, \label{eq:quench}
\end{eqnarray}
where $ \langle r^2 \rangle$ is the mean squared radius of the light nucleus. 

\begin{figure}
\begin{center}
\includegraphics[angle=0,width=0.42\textwidth]{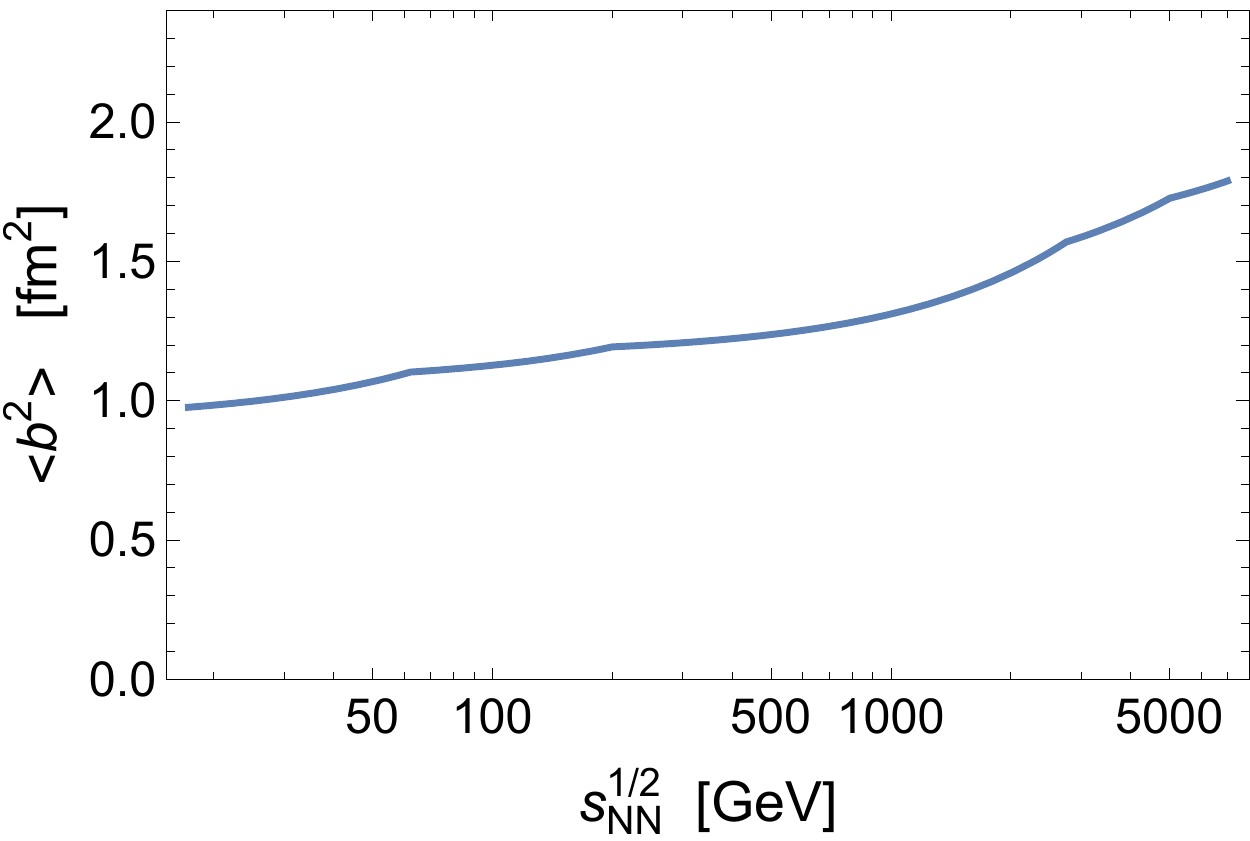}
\end{center}
\vspace{-6mm}
\caption{The dispersion parameter $\langle b^2 \rangle$ due to interactions of the light nucleus with the large nucleus, plotted as a function of the 
collision energy.  \label{fig:b2}}
\end{figure}

The value of $\langle b^2 \rangle$ as a function of the collision energy, obtained from the Gamma wounding profile~\cite{Rybczynski:2013mla}, 
which realistically describes the $pp$ collision data, 
is shown in Fig.~\ref{fig:b2}. Then, for the light nuclei displayed in Table~\ref{tab:Q2}, the quenching factor for the ellipticity is 70-80\%. 

\subsection{Elliptic flow \label{sec:elliflow}}

The harmonic flow vector determined from the one-body azimuthal distribution of particle momenta in a single event, $dN^{\rm ev}/d\phi$, is
\begin{eqnarray}
 \vec{v}_n  &=& v^x_n + i  v^y_n  
 =  \frac{\int  d\phi e^{i n \phi} \frac{dN^{\rm ev}}{d\phi}}{\int  d\phi \frac{dN^{\rm ev}}{d\phi}}, \label{eq:v} 
\end{eqnarray}
where $n$ indicates the Fourier rank. 
To an approximation sufficiently good for our rough estimates, the 
eccentricity and flow vectors are proportional to each other event by event for $n=2$ and $3$. In particular, for the needed $n=2$ case
(ellipticity)
\begin{equation}
\vec{v}_2 \simeq k \vec{\epsilon}_2. \label{eq:prop}
\end{equation}
For the considered small systems and energies, the response coefficient is approximately $k \sim 0.2$~\cite{Nagle:2013lja}. 


Correspondingly, 
$v_2\{\Phi_P \}$  can be determined using  the one-particle distributions, since
\begin{equation}
\frac{dN}{d\phi} \propto 1+ 2v_2\{ \Phi_P \} \cos\left[ 2(\phi-\Phi_P) \right] +\dots ,
\label{eq:azpol}
\end{equation}
with $\Phi_P$ fixed and known.
As discussed in a greater detail in~\cite{Bozek:2018xzy}, this one-body definition has important advantages over the statistical methods needed to extract 
various flow coefficients from correlation
measurements, where non-flow effects pose a limit to the sensitivity of flow measurements at low multiplicites  \cite{Borghini:2001vi}.

\subsection{Imperfect polarization \label{sec:imppol}}

Experimental realizations of our proposal need to utilize polarized targets, where polarization is 
never perfect~\cite{Mane:2005xh,Alekseev:2003sk}.
For particles with $j=1$, 
the tensor polarization, relevant for the case of the deuteron, is defined as 
\begin{eqnarray}
P_{zz}=n(1)+n(-1)-2 n(0),
\end{eqnarray}
where $n(j_3)$ is the fraction of states with angular momentum projection $j_3$. 
Since for $j_3=0$ the magnitude of the eccentricity of the fireball is about twice as large as for $j_3=\pm 1$,
the estimated elliptic flow with respect to the 
polarization axis $\Phi_P$ for partially polarized $j=1$ targets is 
\begin{eqnarray}
v_2\{\Phi_P\} \simeq k\,  \epsilon_2^{j_3=\pm1}\{\Phi_P\} P_{zz}. \label{eq:polex}
\end{eqnarray}
For the deuteron, the experimentally accessible polarization is  $-1.5 \lesssim P_{zz} \lesssim 0.7$~\cite{Sato:1996te,Savin:2014sva}.
Similar formulas can be provided for higher spin states.

\section{Estimate of ellipticity of nuclear distributions from the quadrupole moment \label{sec:est}}

In this Section we show that the ellipticity of the nuclear distribution may be effectively estimated 
from  the nuclear quadrupole moment  $Q_2$ and the mean squared charge radius of the nucleus, which are experimentally well known quantities.
This is convenient from a practical point of view, as there is no need for
numerical simulations to get a rough estimate of $v_2\{\Phi_p\}$.

We use Eq.~(\ref{eq:ratio3}) as the staring point. 
Electromagnetic scattering probes the distributions of charge only, hence we have access to moments of
the distribution of protons
and not all the nucleons, as needed for Eq.~(\ref{eq:ratio3}).
However, for the purpose of our rough estimate we may assume that these distributions are close to each other. 
As a matter of fact, for the case of the deuteron they are 
trivially exactly the same, as $\vec{r}_1=-\vec{r}_2$. For other light nuclei there is admittedly some departure between 
the proton and neutron densities. 
The estimate of the difference in the ms radius $\langle r^2 \rangle$ between neutrons 
and protons due to the neutron skin effect  is  less than 5\% in $^{208}$Pb \cite{Tarbert:2013jze}. The effect on the ratio in Eq. (\ref{eq:epsphiPav}) is probably smaller.

The charge ms radius is defined as 
\begin{eqnarray}
\langle r^2 \rangle_{\rm ch}= \frac{1}{Z} \sum_{i=1}^Z \langle r_i^2 \rangle_{\rm ch} = \frac{1}{Z} \sum_{i=1}^Z \langle x_i^2 +y_i^2+z_i^2 \rangle_{\rm ch}.
\end{eqnarray}
Since the distributions are axially symmetric about the $x$ axis (the polarization axis), we can write 
\begin{eqnarray}
\langle r^2 \rangle_{\rm ch}= \langle x^2 + 2y^2 \rangle_{\rm ch}. \label{eq:charge}
\end{eqnarray}
For strong interactions, pertinent to our study, we need to unfold the proton size effect, which leads to
the ms radius of the distribution of the centers of nucleons, 
\begin{eqnarray}
&& \langle x^2 \rangle =\langle x^2 \rangle_{\rm ch} - \tfrac{1}{3} \langle r^2 \rangle_p, \nonumber \\
&& \langle y^2 \rangle =\langle y^2 \rangle_{\rm ch} - \tfrac{1}{3} \langle r^2 \rangle_p, \nonumber \\
&& \langle r^2 \rangle  = \langle r^2 \rangle_{\rm ch} -  \langle r^2 \rangle_p
\end{eqnarray}

The electric quadrupole moment is defined as
\begin{eqnarray}
Q_2=\sum_{i=1}^Z \left \langle 3x_i^2-r_i^2 \right \rangle \label{eq:Q2def}
\end{eqnarray}
(note the summation and not averaging over the charges). With the above-mentioned symmetry 
\begin{eqnarray}
Q_2=  2Z \langle x^2-y^2 \rangle_{\rm ch}.  \label{eq:Q2a}
\end{eqnarray}
Note that this quantity is not altered by the proton electromagnetic size unfolding.

Relations (\ref{eq:charge}) and (\ref{eq:Q2a}) allow us to estimate 
the ellipticity of the nuclear distribution of Eq.~(\ref{eq:ratio3}) as
\begin{eqnarray}
\epsilon^{|\Psi|^2}_2\{\Phi_P\} &=& - \frac{\langle x^2-y^2 \rangle}{\langle \tfrac{2}{3}(x^2+2y^2)+\tfrac{1}{3}(x^2-y^2) \rangle}  \simeq - \frac{3 Q_2}{4Z \langle r^2 \rangle},
\nonumber \\ \label{eq:Q2}
\end{eqnarray}
where we keep only the leading term in $Q_2$.

Definition (\ref{eq:Q2def}) is equivalent in a standard way to 
\begin{eqnarray}
Q_2 = \langle r^2 \sqrt{\tfrac{16\pi}{5}}Y_{20}(\Omega) \rangle,
\end{eqnarray}
where $Y_{lm}(\Omega)$ denotes the spherical harmonic function.
From the Wigner-Eckart theorem ($\hat{Q}_{20}=r^2 \sqrt{\tfrac{16\pi}{5}}Y_{20}(\Omega)$ is a rank-2 tensor) one has 
\begin{eqnarray}
\langle j j_3 | \hat{Q}_{20} |j j_3 \rangle = \langle j j_3 2 0 | j j_3 \rangle \langle j || \hat{Q}_2 || j \rangle , \label{eq:cg}
\end{eqnarray}
which relates the values of the quadrupole moment for various $j_3$ states by the Clebsch-Gordan coefficients 
(experimentally, the quoted values for $Q_2$ correspond by convention to the highest spin state, $j_3=j$).
Moreover, the lowest possible $j$ to support nonzero $Q_2$ is 1.  
Therefore, the effect discussed in this paper is absent for instance for $^3$He or tritium, where $j=\tfrac{1}{2}$.

The estimates for $\epsilon^{|\Psi|^2}_2\{\Phi_P\}$ following from Eq.~(\ref{eq:Q2}) for several light nuclei are collected in Table~\ref{tab:Q2}.
We note that the expected size of the effect for $^7$Li, $^9$Be or $^{10}$B is of the order of 20\%, significantly larger that for the deuteron (for which the use of the estimate is 
somewhat abusive in view of the discussion at the end of Sect.~\ref{sec:generic}, nevertheless the values are not far from more precise numbers of Eq.~(\ref{eq:deutnum})).

\begin{table}
\caption{Experimental values of the nuclear rms radii 
$\langle r^2 \rangle_{\rm ch}^{1/2}$~\cite{ANGELI201369}, electric quadrupole moments $Q_2$~\cite{STONE200575}
(see also~\cite{048_Mertzimekis}),
and the resulting estimate for the ellipticity of the nuclear distribution from Eq.~(\ref{eq:Q2}) and its leading term $-3Q_2/4Z \langle r^2 \rangle$. \label{tab:Q2}}
\begin{center}
\begin{tabular}{c|rrrrrr}
      & $j$ & $ j_3$      & $\langle r^2 \rangle_{\rm ch}^{1/2}$ [fm] & $Q_2$ [fm$^2$] 
      & $-\frac{3Q_2}{4Z \langle r^2 \rangle}$  [\%] \\ \hline 
 d   & $1$ & $\pm 1$ & 2.1421(88)                                 & 0.2860(15)           & $-5.6$ \\
      &        & $0$        &                                                    & $\times (-2)$        & $\times (-2)$ \\ \hline
 $^7$Li  & $\tfrac{3}{2}$ & $\pm \tfrac{3}{2}$ & 2.444(42)                              & $-4.03(4)$           & $19$  \\
      &        &    $\pm \tfrac{1}{2}$     &                                                & $\times (-1)$        &   $\times (-1)$\\   \hline   
 $^9$Be  & $\tfrac{3}{2}$ & $\pm \tfrac{3}{2}$ & 2.519(12)                     & $5.29(4)$           & $-17$\\
      &        &    $\pm \tfrac{1}{2}$     &                                                & $\times (-1)$       &  $\times (-1)$ \\ \hline
 $^{10}$B  & $\pm 3$               & $\pm 3$                    & 2.428(50)                     & $8.47(6)$             & $-25$\\
       &         & $\pm 2$        &                                                    & $\times 0$        &   0 \\ 
      &        & $\pm 1$        &                                                    & $\times (-3/5)$        & $\times (-3/5)$ \\ 
       &        & $0$        &                                                    & $\times (-4/5)$        & $\times (-4/5)$ \\
\end{tabular}
\end{center}
\end{table}

Joining Eqs.~(\ref{eq:quench},\ref{eq:prop},\ref{eq:Q2}) we find the combined estimate for the elliptic flow coefficient 
evaluated with respect to the polarization axis 
\begin{eqnarray}
v_2\{\Phi_P\} \simeq - k \frac{3 Q_2}{4Z( \langle r^2 \rangle+ \tfrac{3}{2}\langle b^2 \rangle)} \frac{3j_3^2-j(j+1)}{j(2j-1)}, \label{eq:finest}
\end{eqnarray}
where we include the explicit Clebsch-Gordan coefficients from Eq.~(\ref{eq:cg}). The formula holds for 
perfectly polarized light nuclei, central collisions with sufficiently large number of sources, and $j \ge 1$.

\section{Glauber simulations of ellipticity \label{sec:Glauber}}

In the present study we use the wounded nucleon model~\cite{Bialas:1976ed} with a binary component~\cite{Kharzeev:2000ph}, as implemented in
{\tt GLISSANDO}~\cite{Rybczynski:2013yba,Bozek:2019wyr}.
The initial entropy is proportional to 
$S= {\rm const} \left ( N_{\rm W}/2 + a N_{\rm bin} \right )$, 
where $N_{\rm W}$ and   $N_{\rm bin}$
denote the numbers of the wounded nucleons and binary collisions, respectively, and $a=0.145$ for the considered collision energy. 
The entropy produced at the NN collision point in the transverse plane  is 
smeared with a Gaussian of width $0.4$~fm, as is typically done for the initialization of the hydrodynamic studies.

\begin{figure}
\begin{center}
\includegraphics[angle=0,width=0.5\textwidth]{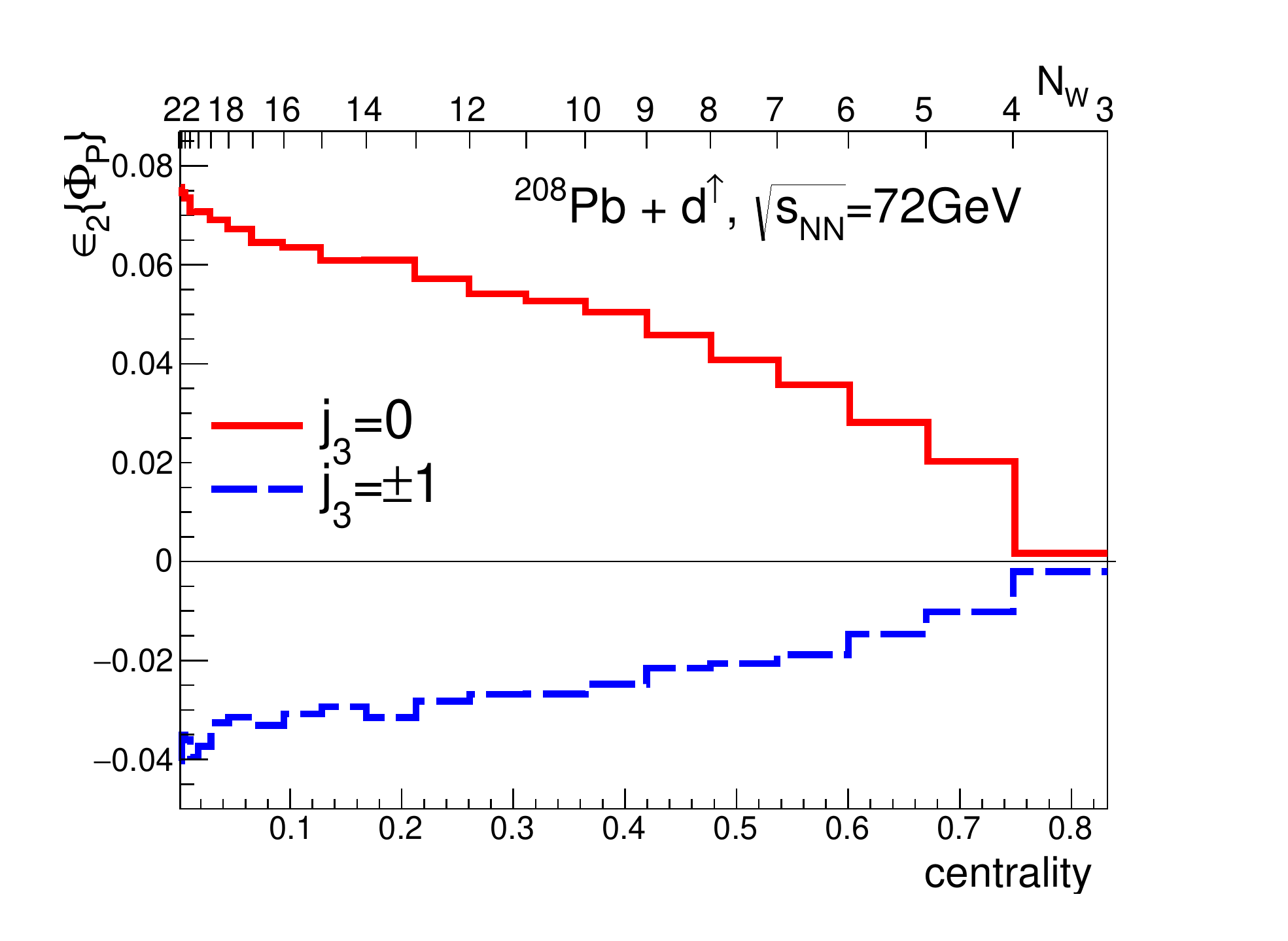}
\end{center}
\vspace{-9mm}
\caption{Ellipticities evaluated in reference to the fixed polarization axis, $\epsilon_2\{\Phi_P\}$, of the fireball formed in Pb collisions on a polarized deuteron at
the collision energy $\sqrt{s_{NN}}=72$~GeV. 
The lower coordinate axis gives the centrality as defined via the initial entropy $S$.
The top coordinate axis is labeled with the corresponding number of the wounded nucleons.  \label{fig:e2P}}
\end{figure}

Our results for  $\epsilon_2\{\Phi_P\}$ of the fireball created in collisions of Pb  on a polarized deuteron target are shown in Fig.~\ref{fig:e2P}.
We have chosen
the collision energy of $\sqrt{s_{NN}}=72$~GeV, which is planned for the future fixed target experiments at the LHC. 
We plot $\epsilon_2\{\Phi_P\}$ as a function of the centrality of the collision
defined via quantiles of the distribution of the initial entropy $S$. 
For the reader's convenience, we also give the corresponding number of the
the wounded nucleons, $N_{\rm W}$, along the top coordinate axis. 
For the most central collisions, the ellipticities 
of the fireball are about $\sim 50\%$ smaller compared to the ellipticities of the distributions of the polarized deuteron. 
This reduction is caused by the contribution from the Pb nucleons, whose positions fluctuate randomly. 
From geometric arguments, the magnitude of $\epsilon_2\{\Phi_P\}$ drops to zero for peripheral collisions. 
We note that the relation $\sum_{j_3} \epsilon_2^{j_3}\{\Phi_P\} \simeq 0$ is satisfied numerically, in agreement to 
the corresponding relation for the eccentricities of the deuteron nuclear distributions.

\begin{figure}
\begin{center}
\includegraphics[angle=0,width=0.5\textwidth]{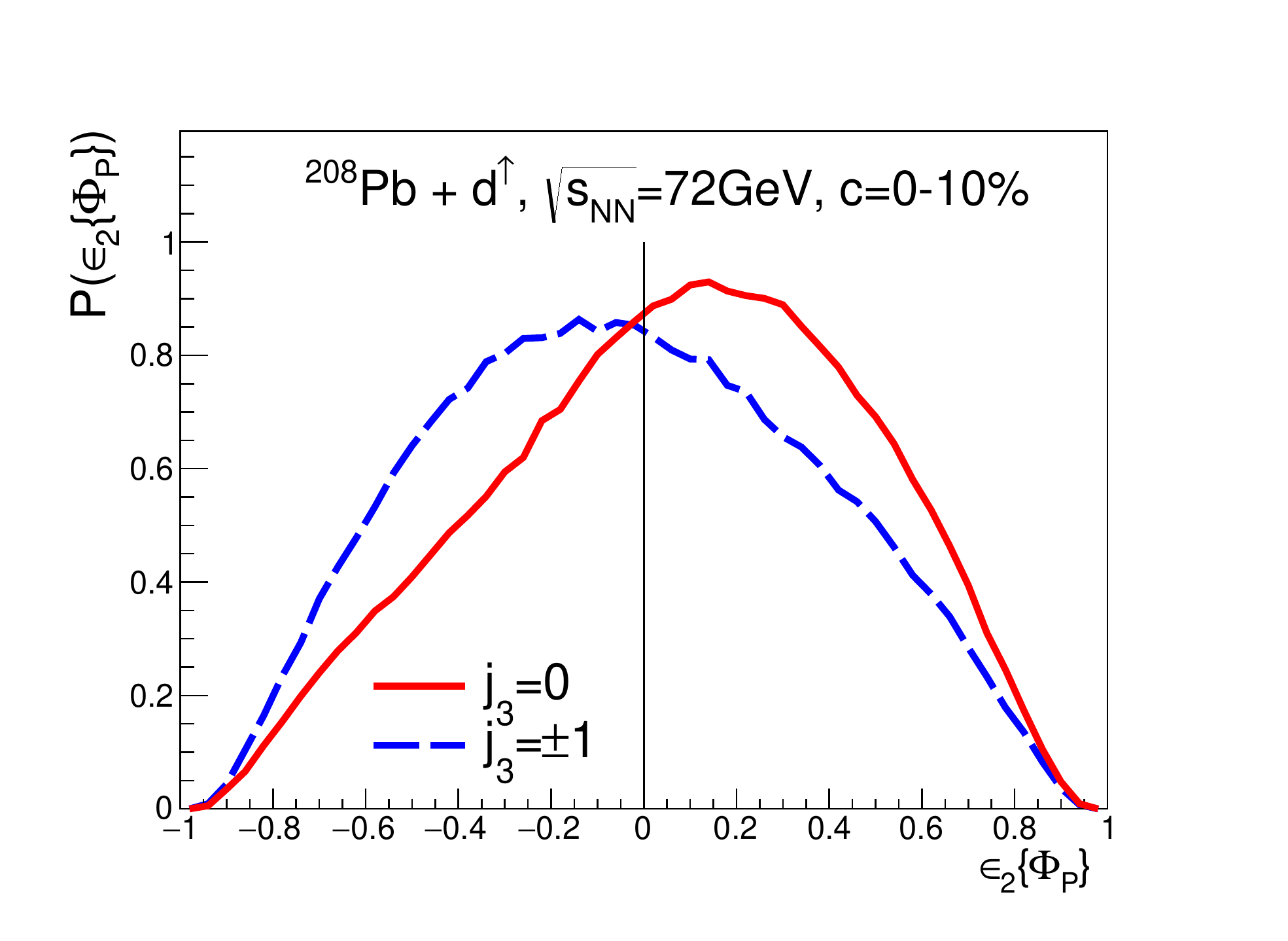}
\end{center}
\vspace{-9mm}
\caption{Distribution of ellipticity $\epsilon_2\{\Phi_P\}$ of the fireball formed in Pb collisions on a polarized deuteron 
at $\sqrt{s_{NN}}=72$~GeV and centrality $c=0-10\%$. \label{fig:epsdist}}
\end{figure}

The size of $\epsilon_2\{\Phi_P\}$ is at the level of a few percent. Using Eq.~(\ref{eq:polex}) and the range of $P_{zz}$ for the deuteron 
yields the estimate for the flow coefficient in most central collisions as
\begin{eqnarray}
-0.5\% \lesssim v_2\{\Phi_P\} \lesssim 1\%.
\end{eqnarray}
As this quantity is measured in reference to the zero result (which would be the case in the absence of polarization or collective evolution), it 
should be easily accessible to future experiments at the typical statistics accumulated in heavy-ion collisions. 

The event-by-event distribution of  $\epsilon_2\{\Phi_P\}$ for the most central collisions ($c=0-10\%$) is provided in Fig.~\ref{fig:epsdist}. We note a visible shift of the distribution 
towards positive values for the $j_3=0$ polarization, and in the opposite direction for $j_3=\pm 1$, which corresponds to the mean values plotted in  Fig.~\ref{fig:e2P}.

\begin{figure}
\centering
\includegraphics[angle=0,width=0.5 \textwidth]{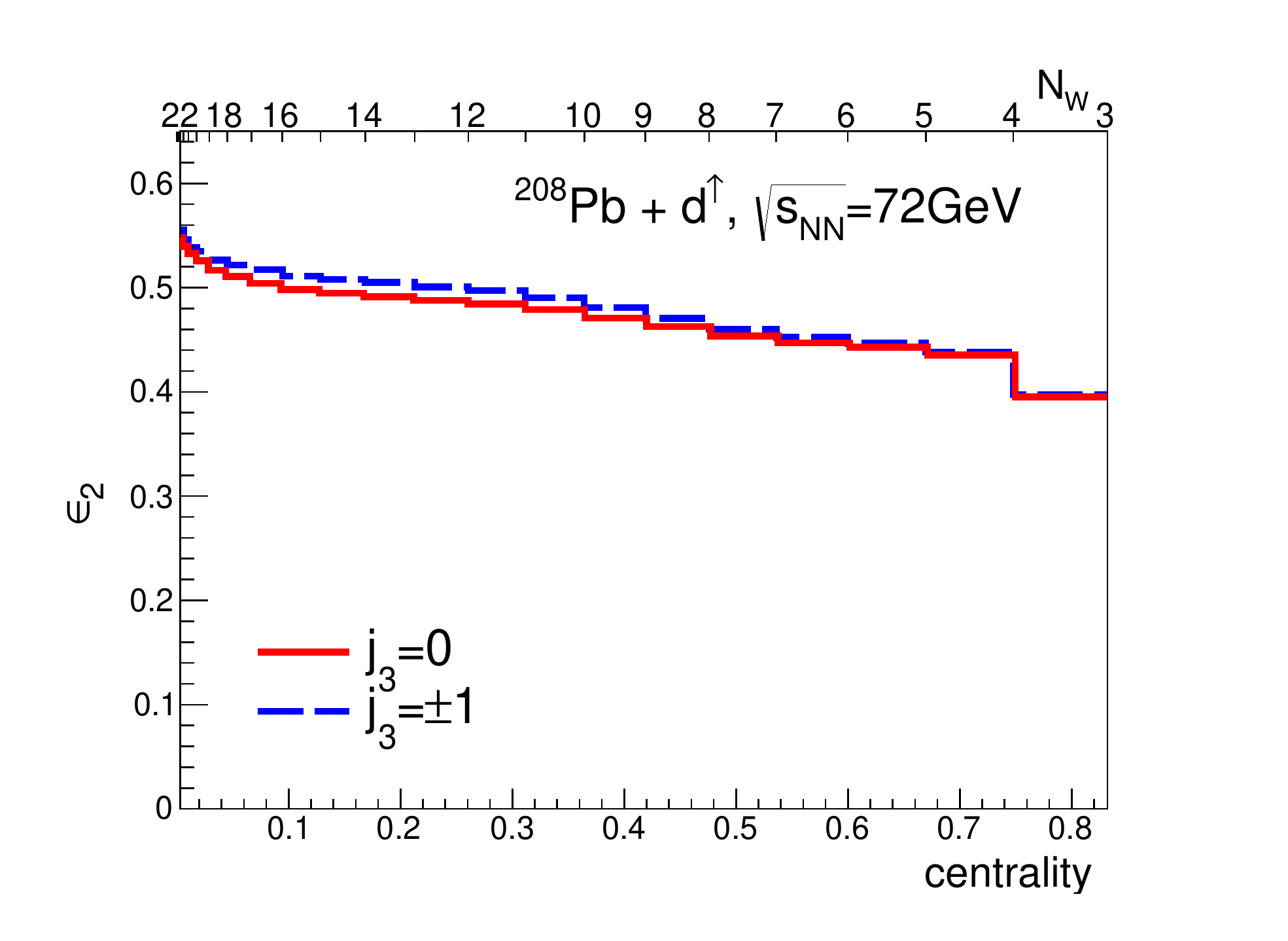}
\vspace{-9mm}
\caption{Same as in Fig.~\ref{fig:e2P} but for the participant-plane ellipticity $\epsilon_2$. 
It is dominated by fluctuations and the relative splitting effect between the $j_3=0$ and $j_3=\pm 1$ cases is small. \label{fig:e2}}
\end{figure}

In Fig.~\ref{fig:e2} we show, by contrast to Fig.~\ref{fig:e2P}, 
the participant-plane ellipticity $\epsilon_2 = |\vec{\epsilon}_2|$. We note that the relative difference between the $j_3=0$ and $j_3=\pm 1$ cases is 
tiny, making the possible experimental resolution between the two polarizations very difficult when using standard  two-particle correlation measures for $v_2$.

\begin{figure}
\begin{center}
\includegraphics[angle=0,width=0.5\textwidth]{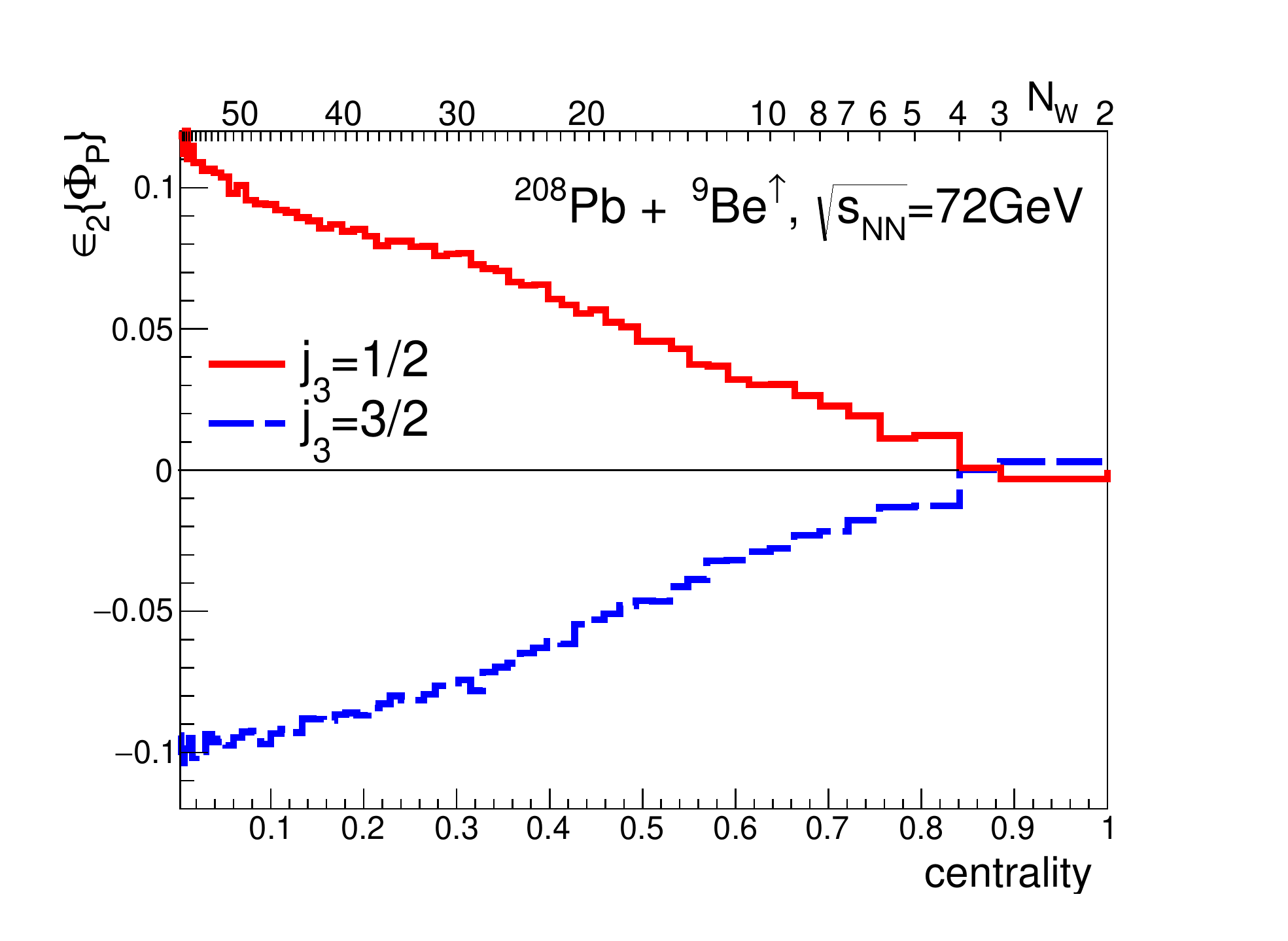}
\end{center}
\vspace{-9mm}
\caption{Same as in Fig.~\ref{fig:e2P} but for Pb collisions on polarized ${}^9$Be target. \label{fig:9Be}}
\end{figure}

Finally, in Fig.~\ref{fig:9Be} we present an analogous study to Fig.~\ref{fig:e2P} but for Pb collisions on a polarized ${}^9$Be target. We expect, according to the estimates from 
Table~\ref{tab:Q2} a larger effect than for the deuteron, which is indeed the case. Here the {\tt GLISSANDO} simulations use the clustered ${}^9$Be
distributions as described in~\cite{Rybczynski:2017nrx}.

\section{Experimental prospects and further outlook \label{sec:outlook}}

Future fixed target experiments at the LHC (AFTER@LHC), and in particular SMOG2@LHCb)~\cite{Barschel:2014iua,Aaij:2014ida}, 
plan to study collisions of a 2.76A~TeV Pb beam on a fixed target. This collision energy corresponds to 
$\sqrt{s_{NN}}=72$~GeV, which falls between CERN SPS and top BNL RHIC energies. 
The rapidity coverage of a fixed target experiment is shifted, as $y_{\rm CM} = 0$  corresponds to  $y_{\rm lab}=4.3$.
As the LHCb detector has the pseudorapidity coverage $2<\eta<5$, in the NN CM frame of a fixed target experiment 
it would correspond to $-2.3<\eta<0.7$, which is the midrapidity region intensely studied in other ultra-relativistic collisions 
up to now. There are possibilities of using polarized targets through
available standard technology and, hopefully, such efforts will be undertaken.

We note that the effect of non-zero elliptic flow evaluated in reference to the polarization axis occurs for nuclei with 
angular momentum $j\ge 1$, and can be estimated based on their ms radius and  quadrupole moments.  
As shown in this work, other nuclei in addition to the deuteron,  such as ${}^7$Li, ${}^9$Be, or ${}^{10}$B, are even better for such studies.
If the effect is indeed confirmed, it would make another strong case for a late stage generation of collectivity seen in light-heavy ultra-relativistic nuclear collisions.

Other opportunities emerging in collision with light polarized targets are worth mentioning, such as studies of hard probes (jets,
photons, heavy flavor mesons) relative to the polarization axis $\Phi_P$, or interferometry correlations defined relative to $\Phi_P$.

\begin{acknowledgments}
We thank Thomas Shaffer for suggesting a relation of eccentricity to the quadrupole moment and Adam Bzdak for discussions on alternative
explanations of collectivity in small systems, and Wojciech Florkowski for remarks concerning polarization.
Project supported by the  Polish National Science Centre grants 2015/19/B/ST2/00937 (WB) and 2018/29/B/ST2/00244 (PB).
\end{acknowledgments} 

\appendix

\section{The deuteron wave function \label{app:deut}}

\begin{figure}[tb]
\centering
\includegraphics[angle=0,width=0.37 \textwidth]{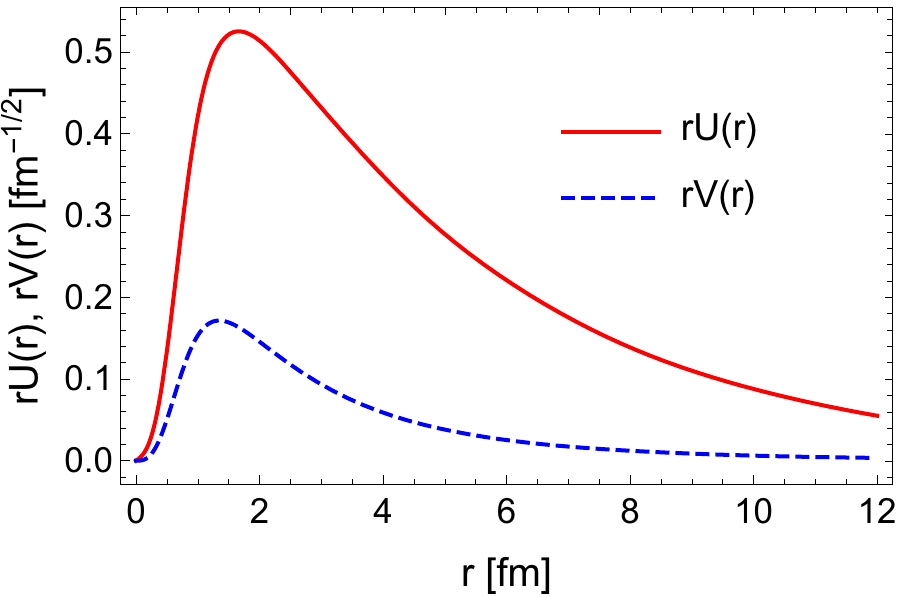}
\vspace{-2mm}
\caption{Radial wave functions of the $S$-wave, $U(r)$, and $D$-wave, $V(r)$, components of the deuteron, multiplied by the relative radius 
$r$, taken from the parametrization provided in~\cite{Zhaba:2015yxq} for  Reid93 nucleon-nucleon potential.  \label{fig:reid93}}
\end{figure}

In this Appendix we give for completeness the standard features of the deuteron wave function. 
In the ground state, the deuteron has $j^P=1^+$ quantum numbers, with a dominant ${}^3S_{1}$-wave contribution and a 
small ${}^3D_{1}$-wave admixture. The state with $j_3$ projection of the total angular momentum $j$ is
\begin{eqnarray}
| \Psi(r;j_3) \rangle &=& U(r)|j=1,j_3,L=0,S=1 \rangle  \nonumber \\ &+&  V(r)|j=1,j_3,L=2,S=1 \rangle,  
\end{eqnarray}
where $r$ in the relative radial coordinate of the p-n separation, 
and $U(r)$ and $V(r)$ denote the $S$- and $D$-wave radial wave functions, respectively.  
The Clebsch-Gordan decomposition into the orbital momentum and spin states, $|LL_3\rangle |SS_3\rangle$, yields 
\begin{eqnarray}
&& | \Psi(r;1)\rangle = U(r) |00 \rangle |11 \rangle \label{eq:02} \\ 
&& ~+ V(r)  \Big [ \sqrt{\tfrac{3}{5}}  |22 \rangle |1-\!\!1 \rangle -  \sqrt{\tfrac{3}{10}}  |21 \rangle |10 \rangle + \sqrt{\tfrac{1}{10}}  |20 \rangle |11 \rangle \Big ] , \nonumber \\
&& | \Psi(r;0)\rangle = U(r) |00 \rangle |10 \rangle \nonumber  \\ 
&& ~+ V(r)  \Big [ \sqrt{\tfrac{3}{10}}  |21 \rangle |1-\!\!1 \rangle -  \sqrt{\tfrac{2}{5}}  |20 \rangle |10 \rangle + \sqrt{\tfrac{3}{10}}  |2-\!\!1 \rangle |11 \rangle \Big ] . \nonumber
\end{eqnarray}
Orthonormality of the spin components gives immediately the probability distributions 
\begin{eqnarray}
&& | \Psi(r,\theta,\phi;\pm 1)|^2 = \frac{1}{16\pi}  \left [ 4 U(r)^2 - \right . \label{eq:dens} \\ 
&& ~~\left . 2 \sqrt{2} \left(1-3 \cos ^2(\theta ) \right) U(r) V(r)+\left(5-3 \cos ^2(\theta )\right) V(r)^2 \right ], \nonumber \\
&& | \Psi(r,\theta,\phi;0)|^2 = \frac{1}{8\pi}  \left [ 2 U(r)^2 + \right . \nonumber \\ 
&& ~~\left . 2 \sqrt{2} \left(1-3 \cos ^2(\theta )\right) U(r) V(r)+\left(1+3 \cos ^2(\theta )\right) V(r)^2 \right ],  \nonumber
\end{eqnarray}
with 
$\sum_{j_3} | \Psi(r,\theta,\phi;j_3)|^2=\frac{3}{4\pi}[U(r)^2+V(r)^2]$. We use the normalization 
$\int r^2 dr (U(r)^2+V(r)^2)=1$.

Several features are worth stressing. First, because $V(r)^2 \ll U(r)^2$, the terms proportional to $U(r) V(r)$ in Eq.~(\ref{eq:dens})
stemming from the interference term of the spin $|11\rangle$ 
components in  Eq.~(\ref{eq:02}), are responsible for a significant polar angle dependence, whereas 
the terms proportional to $V(r)^2$ are negligible. 
Second, the distributions are oblate for 
$j_3=0$ and prolate for $j_3=\pm1$  (cf. Fig.~\ref{fig:polar}).

Numerous parameterizations of the deuteron wave functions are available in the literature~\cite{Zhaba:2015yxq}, leading to similar results. 
For the estimates provided in Sec.~\ref{sec:est} we use the 
deuteron wave functions from Reid93 nucleon-nucleon potential, presented in Fig.~\ref{fig:reid93}. 
In Reid93 parametrization, the weight of the $D$-wave component is
$\int_0^\infty V(r)^2 r^2 dr = 5.7\%$, clearly showing the strong $S$-wave dominance.

\section{Ellipticity of the deuteron distribution \label{app:eldeu}}

We apply the generic definition of ellipticity from Eq.~(\ref{eq:epsphiPav}) to the distribution of nucleons in the 
projectile deuteron. Passing to relative spherical coordinates, we find
\begin{eqnarray}
\epsilon_2^{|\Psi|^2_{j_3}}\{\Phi_P\} &=& 
 \int_0^\infty  \!\!\!\! r^2  dr \!  \int \!d\Omega\, \left | \Psi(r,\theta,\phi;j_3) \right |^2   \nonumber \\
&\times& \frac{\cos^2\theta -\sin^2 \theta \cos^2\phi}{\cos^2\theta -\sin^2 \theta \cos^2\phi}, \label{eq:ratio}
\end{eqnarray}
which upon explicit evaluation with the wave functions (\ref{eq:dens}) yields 
\begin{eqnarray}
\epsilon_2^{|\Psi|^2_{j_3=0}}\{\Phi_P\} &=& \frac{1}{4} \int r^2 dr \left [ 2 \sqrt{2} U(r)V(r) -V(r)^2 \right], \nonumber \\
\epsilon_2^{|\Psi|^2_{j_3=1}}\{\Phi_P\} &=& -\frac{1}{2}\epsilon_2^{|\Psi|^2_{j_3=0}}\{\Phi_P\}. \label{eq:ratio12}
\end{eqnarray}
With Reid93 wave functions we find the numbers listed in Eq.~(\ref{eq:deutnum}).
We note that the mixing term, containing $U(r)V(r)$, largely dominates over the $V(r)^2$ term in Eq.~(\ref{eq:ratio12}). 
Obviously, with unpolarized wave function the ellipticity vanishes, as 
$\sum_{j_3=0,\pm 1}\epsilon_2^{|\Psi|^2_{j_3}}\{\Phi_P\}=0$.

\bibliography{hydr}

\end{document}